\documentclass[prm,twocolumn,showkeys,preprintnumbers,amsmath,amssymb,superscriptaddress,aps,10pt]{revtex4-1}

\usepackage{graphicx}
\usepackage{dcolumn}
\usepackage{bm}
\usepackage{amsthm}
\usepackage{amsmath}
\usepackage{amssymb}
\usepackage{epstopdf}

\begin{document}

\title{Towards understanding of magnetization reversal in Nd$-$Fe$-$B nanocomposites: Analysis by high-throughput micromagnetic simulations}

\author{Sergey~Erokhin}\email{s.erokhin@general-numerics-rl.de}
\affiliation{General Numerics Research Lab, Moritz-von-Rohr-Stra{\ss}e~1A, D-07745 Jena, Germany}
\author{Dmitry~Berkov}
\affiliation{General Numerics Research Lab, Moritz-von-Rohr-Stra{\ss}e~1A, D-07745 Jena, Germany}
\author{Masaaki Ito}
\affiliation{Advanced Material Engineering Division, Toyota Motor Corporation, Susono 410-1193, Japan}
\author{Akira Kato}
\affiliation{Advanced Material Engineering Division, Toyota Motor Corporation, Susono 410-1193, Japan}
\author{Masao Yano}
\affiliation{Advanced Material Engineering Division, Toyota Motor Corporation, Susono 410-1193, Japan}
\author{Andreas~Michels}
\affiliation{Physics and Materials Science Research Unit, University of Luxembourg, 162A~Avenue de la Fa\"iencerie, L-1511 Luxembourg, Grand Duchy of Luxembourg}


\begin{abstract}
We demonstrate how micromagnetic simulations can be employed in order to characterize and analyze the magnetic microstructure of nanocomposites. For the example of nanocrystalline Nd$-$Fe$-$B, which is a potential material for future permanent-magnet applications, we have compared three different models for the micromagnetic analysis of this material class: (i) a description of the nanocomposite microstructure in terms of Stoner-Wohlfarth particles with and without the magnetodipolar interaction; (ii) a model based on the core-shell representation of the nanograins; (iii) the latter model including a contribution of superparamagnetic clusters. The relevant parameter spaces have been systematically scanned with the aim to establish which micromagnetic approach can most adequately describe experimental data for this material. According to our results, only the last, most sophisticated model is able to provide an excellent agreement with the measured hysteresis loop. The presented methodology is generally applicable to multiphase magnetic nanocomposites and it highligths the complex interrelationship between the microstructure, magnetic interactions, and the macroscopic magnetic properties.
\end{abstract}

\maketitle\

\section{Introduction}

Nd$-$Fe$-$B based nanocomposite materials for permanent magnets are currently the subject of intensive research efforts due to their promising magnetic properties such as high remanence and large magnetic energy product, which render them attractive for potential applications in electronic devices, motors, and other numerous applications \cite{coehoorn88,Davies1989,gutfleisch2002,gutfleisch2011,liu2009,bance2014}. These materials consist of exchange-coupled nanocrystalline hard ($\mathrm{Nd}_2\mathrm{Fe}_{14}\mathrm{B}$) and soft ($\alpha$-$\mathrm{Fe}$ or $\mathrm{Fe}_3\mathrm{B}$) magnetic phases. The major challenge is the understanding of how the features of the microstructure (e.g., average $\mathrm{Nd}_2\mathrm{Fe}_{14}\mathrm{B}$ particle size and shape, volume fraction of the soft phase, composite texture, interfacial chemical composition) correlate with their macroscopic magnetic properties. This task poses enormous demands both on state-of-the-art experimental methods---such as high-resolution electron microscopy, electron backscattering diffraction, three-dimensional atom-probe analysis, Lorentz and Kerr microscopy, small-angle neutron scattering---and on atomistic and mesoscopic micromagnetic simulations of these materials.

Recent experimental and theoretical studies (see, e.g., Refs.~\onlinecite{liu2013,SepehriAmin2013,liu2014,hrkac2014,fujisaki2014}) have focused on the role of the interface regions separating the $\mathrm{Nd}_2\mathrm{Fe}_{14}\mathrm{B}$ grains. These intergranular regions play an important role for the coercivity of the material, since the magnetic parameters (saturation magnetization, exchange stiffness, magnetocrystalline anisotropy) of grain-boundary layers are different as compared to the $\mathrm{Nd}_2\mathrm{Fe}_{14}\mathrm{B}$ crystallites. Hence the intergrain boundary regions  represent potential sources for the nucleation of inhomogeneous spin textures during magnetization reversal. Besides the different magnetic parameters, the grain-boundaries in Nd$-$Fe$-$B based nanocomposites (with a thickness of about $1-15 \, \mathrm{nm}$) can exist both in a crystalline and in an amorphous state \cite{SepehriAmin2013}.

In order to reveal the relationship between the microstructure and macroscopic magnetic properties, it is important to have a method at hand which allows for the fast and efficient scanning of the multidimensional parameter space required to characterize magnetic nanocomposites. Micromagnetic computations---mainly adopting the meanwhile standard finite-element approach based on tetrahedral finite elements---have been employed for understanding, for example, the coercivity mechanism in Nd$-$Fe$-$B magnets \cite{schrefl94,fischer96,fischer98,fukunaga2001,Gao2003,He2012,Saiden2014,hrkac2014,fujisaki2014,Yi2016}. In this paper, we report on high-throughput simulations of the magnetic microstructure of nanocrystalline Nd$-$Fe$-$B using our new polyhedron-based micromagnetic paradigm, which combines the advantage of a highly flexible meshing with the speed of an FFT-based evaluation of the magnetodipolar field \cite{erokhin2012prb,michels2012prb1,michels2014jmmm,erokhin2015}. Simulation results are compared to an experimental hysteresis loop of a $\mathrm{Nd}_2\mathrm{Fe}_{14}\mathrm{B} / \alpha$-$\mathrm{Fe}$ nanocomposite. We demonstrate that a model where the Nd$-$Fe$-$B grains are represented as Stoner-Wohlfarth (SW) particles---even taking into account the magnetodipolar interaction and the interparticle exchange coupling---cannot satisfactorily explain the measured hysteresis loop. We show that in order to adequately understand the observed hysteresis, we have to invoke a core-shell model of Nd$-$Fe$-$B grains (with reduced values of the anisotropy constant and exchange coupling within the shells and between different grains) supplemented by a Langevin-type (superparamagnetic) high-field contribution.

\section{Methods \label{methods}}

\subsection{Sample under study \label{sample}}

The $\mathrm{Nd}_2\mathrm{Fe}_{14}\mathrm{B}/\alpha$-$\mathrm{Fe}$ nanocomposite (containing $5 \, \mathrm{wt.~\%}$ of $\alpha$-$\mathrm{Fe}$) was prepared by means of the melt-spinning technique. Sample characterization was carried out using transmission electron microscopy and synchrotron x-ray scattering (see Ref.~\onlinecite{michelsPRApplied2017} for details). The melt-spun sample had an average $\mathrm{Nd}_2\mathrm{Fe}_{14}\mathrm{B}$ grain size of about $20 \, \mathrm{nm}$; it represents a nanocomposite material with $\mathrm{Nd}_2\mathrm{Fe}_{14}\mathrm{B}$ as the hard magnetic phase and $\alpha$-$\mathrm{Fe}$ as the soft phase. Magnetization data (up to $\mu_0 H_{\mathrm{max}} = 14 \, \mathrm{T}$) were recorded at $300 \, \mathrm{K}$ using a Cryogenic vibrating sample magnetometer.

\subsection{Micromagnetic simulation methodology \label{mumagmethod}}

Our micromagnetic algorithm was originally developed for the simulation of the magnetization distribution of magnetic nanocomposites and the computation of the related magnetic-field-dependent neutron scattering cross sections (see Refs.~\onlinecite{erokhin2012prb,michels2012prb1,michels2014jmmm,erokhin2015} for details). The four standard contributions to the total magnetic energy (external field, magnetic anisotropy, exchange, and dipolar interaction) are taken into account. In the present study, we have employed two different micromagnetic models, as will be motivated below: the first model was based on the assumption that Nd$-$Fe$-$B grains inside the sample can be represented as uniformly-magnetized Stoner-Wohlfarth (SW) particles; the second model was based on a core-shell-type description of the Nd$-$Fe$-$B grains. Further, the second model has been extended by taking into account the possibility of a superparamagnetic contribution (arising from small magnetic clusters inside the sample) to the measured hysteresis loop. 

For the SW model, the simulation volume was $1.4 \times 1.4 \times 1.4 \, \mathrm{\mu m}^3$, discretized into $4 \times 10^5$ mesh elements. Each mesh element---a polyhedron with a size of $\sim 20 \, \mathrm{nm}$---was supposed to represent one Nd$-$Fe$-$B crystallite. The influence of the soft phase ($\alpha$-Fe) was neglected in this model due to its very low volume fraction ($\sim 5 \, \%$) in the experimentally studied sample. In the simulations of the core-shell model, we had to employ a finer discretization of the simulated sample, because of the more complex structure of the crystallites. This finer discretization has led to an increased computational time, so that the number of mesh elements for this model was limited to $2 \times 10^5$. For this model we could simulate the magnetization reversal of a system consisting of $260$ core-shell nanograins (sample volume: $115 \times 115 \times 115 \, \mathrm{nm}$). The shape of the grains' core was spherical; the core volume fraction could be varied in order to study the effect of this parameter on the hysteresis loop. However, for the majority of simulations, the core volume fraction was fixed at $40 \, \%$, which corresponds to a shell thickness of $\sim 2.6 \, \mathrm{nm}$ (equal to $2$ elementary cells of the Nd$-$Fe$-$B crystal lattice) for an average grain size of $20 \, \mathrm{nm}$. 

Periodic boundary conditions were applied in all simulations. Material parameters used correspond to the standard values of bulk $\mathrm{Nd}_2\mathrm{Fe}_{14}\mathrm{B}$ (see below for further details): saturation magnetization $M_{\mathrm{S}} = 1300 \, \mathrm{G}$, uniaxial magnetocrystalline anisotropy with $K_{\mathrm{bulk}} = 4.3 \times 10^7 \, \mathrm{erg/cm^3}$, and exchange-stiffness constant of $A_{\mathrm{bulk}} = 1.25 \times 10^{-6} \, \mathrm{erg/cm}$. As we will discuss below, these parameters can take on different values in the shell regions. The direction of the anisotropy axis varies randomly from crystallite to crystallite in both models; for the core-shell structure, the directions of the anisotropy axes of both the core and the shell are the same for each individual grain. Throughout the paper, the quality of the fitting is estimated by the normalized difference between experimental and simulated hysteresis loops,
\begin{equation}
\label{fitcriterion}
\Delta = \frac{1}{N_{\mathrm{H}}} \sum_{i=1}^{N_{\mathrm{H}}} \left| M^{\mathrm{exp}}_i - M^{\mathrm{sim}}_i \right| , 
\end{equation}
where $N_{\mathrm{H}}$ is the total number of simulated field values, and $M^{\mathrm{exp}}$ ($M^{\mathrm{sim}}$) represent the total measured (simulated) magnetization projections on the initial applied field direction.

\section{Results and Discussion \label{results}}

\subsection{Model based on Stoner-Wohlfarth particles \label{StonerWohlfarth}}

Taking into account the large magnetocrystalline anisotropy of Nd$-$Fe$-$B and the small size of its grains in the experimental sample ($\approx 20 \, \rm{nm}$), one would expect that the individual grains of this sample remain in a single-domain state during the magnetization-reversal process. We emphasize that the often used estimation of the single-domain critical size $a_{\rm cr}$, based on the comparison of the domain wall (DW) width $\delta_{\rm dw}$ and the particle size $a$ is not applicable here. We remind that this ``criterion'' ($a_{\rm cr} \sim \delta_{\rm dw}$) leads to the statement that the particle remains single-domain as long as its size is smaller than one of the two characteristic micromagnetic lengths: the demagnetizing (magnetostatic) length $l_{\rm dem} \sim (A/M_{\rm S})^{1/2}$ or the exchange length $l_{\rm exch} \sim (A/K)^{1/2}$. The second length for Nd$-$Fe$-$B is very small due to its very large anisotropy constant: $l_{\rm exch}^{\rm Nd-Fe-B} \approx 1.7 \, {\rm nm}$, so that a (Landau-Bloch) DW would easily ``fit'' in a $20$-nm-large Nd$-$Fe$-$B grain. 

However, the domain wall energy (per unit area) $\gamma_{\rm dw} = 4 \sqrt{A K}$ is very high for Nd$-$Fe$-$B due to the large anisotropy constant of this material. Hence, we need a more rigorous estimation of the critical size $a_{\rm cr}$. This estimation can be done based on standard energy arguments: the energy of the multi-domain state must be smaller than that of the single-domain state. For Nd$-$Fe$-$B, its large anisotropy allows one to neglect the magnetodipolar energy contribution, greatly simplifying the estimation. Further, we assume that inside the domains in a Nd$-$Fe$-$B particle the magnetization is directed approximately along the particle anisotropy axis, so that we have to compare contributions from the external field energy and the DW energy only. For a rough estimation of $a_{\rm cr}$, we consider the situation in a negative external field (i.e., the field is directed opposite to the initial saturation direction). In this case the energy difference $\Delta E$ between the state saturated in the initial field direction and the state containing a domain with a characteristic size $a$, surface area $S_{dw}$, and volume $V_{\rm nuc}$ is
\begin{equation}
\label{DeltaE}
\begin{aligned}
\Delta E = {} & -2 M_{\rm S} H_{\rm ext} V_{\rm nuc} + \gamma_{\rm dw} S_{dw} = \\
&  -2 c_v a^3 M_{\rm S} H_{\rm ext} + c_s a^2 4 \sqrt{A K},
\end{aligned}
\end{equation}
where we have used the above mentioned expression for the DW energy $\gamma_{\rm dw}$ and introduced the proportionality coefficients $c_v$ ($c_s$) between the domain volume (surface area) and the corresponding powers of the domain size $a$ as $V_{\rm nuc} = c_v a^3$ and $S_{dw} = c_s a^2$. The critical single-domain size deduced from the statement that the particle remains single-domain as long as $\Delta E \ge 0$ leads to the following critical size:
\begin{equation}
\label{CritSize1}
a_{\rm cr} = 2 \frac{c_s}{c_v} \frac{\sqrt{A K}}{M_{\rm S} H_{\rm ext}}
\end{equation}
We note that the critical size obtained in this way depends not only on the magnetic material parameters and the domain shape (via the relation $c_s/c_v$), but also on the external field value $H_{\rm ext}$. This is a natural consequence of the fact that the energy competing with the domain wall energy is in our case the energy due to the external field.

For a domain {\it inside} the bulk of the material, we have $c_s/c_v = 6$ both for a spherical and a cubical domain. Substituting into Eq.~(\ref{CritSize1}) the standard materials parameters of Nd$-$Fe$-$B and the experimentally found coercivity $H_c \approx 6 \, {\rm kOe}$ for the external field value $H_{\rm ext}$, we obtain $a_{\rm cr} \approx 110 \, {\rm nm}$. This critical size is far above the particle diameter of the here studied Nd$-$Fe$-$B sample. An energetically more favorable configuration is the formation of a domain as a spherical segment with the height $a$ near the {\it surface} of the particle with radius $R$. In this case the estimation of $a_{\rm cr}$ can also be cast into the form of Eq.~(\ref{CritSize1}) with the geometrical factor $c_s/c_v$ varying between 1.5 and 2.0, depending on the relation between $a$ and $R$. For such a domain, we obtain in the ``best'' case the value $a_{\rm cr} \approx 30 \, {\rm nm}$, which is still considerably larger than the experimental particle size.

It is important to understand that the estimation above is based on the comparison of system energies in the initial (single-domain) and final (multi-domain) states and thus provides only the critical size of the nucleation region in complete equilibrium (we note in passing that in order to make this estimation more rigorous, the entropic contribution must be added, which should be evaluated for each particular configuration and most probably will not significantly change the final result). By studying real magnetization reversal, the height of the energy barrier between the initial and final states should be taken into account, in order to be able to estimate the transition time between the two states. In our case, this transition corresponds to the path over the anisotropy energy barrier by the magnetization rotation inside the nucleation region. The height of this barrier $E_{\rm anis} = K V$ is (for Nd$-$Fe$-$B) much higher than the thermal energy $k T$ already for very small nucleation region sizes: for a spherical nucleus $E_{\rm anis} = 100 \, k T$ for $a \approx 5.7 \, {\rm nm}$ and at $T = 300 \, \rm K$. Hence, for sizes larger than a few nanometers, a corresponding transition is not possible within a realistic time scale. This argument additionally supports the expectation that magnetization reversal of a Nd$-$Fe$-$B particle with a diameter of about $20 \, {\rm nm}$ and an intact crystal structure should be well described by the SW model. 

Therefore, our first model of the $\mathrm{Nd}_2\mathrm{Fe}_{14}\mathrm{B}$ nanocomposite is based on the representation of the sample as an ensemble of SW particles, i.e., homogeneously magnetized particles having uniaxial magnetic anisotropy. Between these particles both the magnetodipolar and the exchange interactions may be present. In order to study the influence of different magnetic parameters on the magnetization process separately, we have run a large set of micromagnetic simulations where both the anisotropy constant of the particles and the exchange-coupling constant between them have been varied. The physical reasons behind this approach were the following: (i) the anisotropy constant of nanosized magnetic particles could significantly deviate form its bulk value due to the disturbed crystal structure of such particles (especially near their surface) and (ii) the interparticle exchange interaction can be arbitrary small due to the poor quality of intergrain boundaries in such materials. From the methodological point of view, it was important to find out whether the collection of nanosized Nd$-$Fe$-$B particles obtained via the melt-spinning technique can be described by the SW model, being extended to include interparticle interactions.

\begin{figure}[htb]
\centering
\resizebox{0.7\columnwidth}{!}{\includegraphics{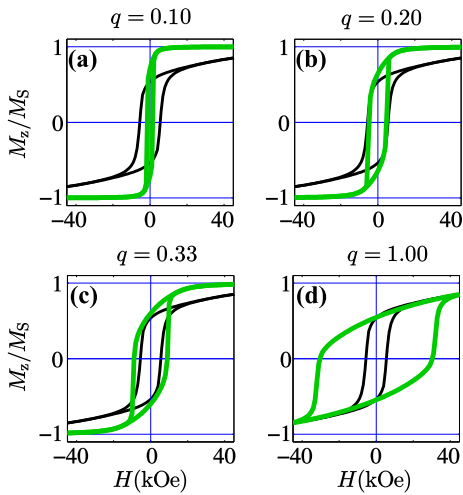}}
\caption{Simulated hysteresis loops (green) obtained for the model based on Stoner-Wohlfarth particles with magnetodipolar interaction for different anisotropy constants (exchange-coupling constant $\kappa = 0.33$ for all loops). Black curve on each graph is the experimentally measured loop.}
\label{HystSWwithdipoleKAPPA033}
\end{figure}
Figure ~\ref{HystSWwithdipoleKAPPA033} displays a subset of results of these simulations, where the magnetodipolar interparticle interaction and exchange interaction are taken into account. The more complete ``matrices'' of hysteresis loops in the coordinates ``exchange constant--anisotropy constant'' are presented in the Supplemental Material. In Fig.~\ref{HystSWwithdipoleKAPPA033} the values of the grain anisotropy constant $K$ and the intergrain exchange constant $A$ are shown on each plot as the values of the reduced constants $q$ and $\kappa$, which are defined via the corresponding bulk parameters as $K = q K_{\mathrm{bulk}}$ and $A = \kappa A_{\mathrm{bulk}}$ (so that $0 \leq q \leq 1$ and $0 \leq \kappa \leq 1$). 

This figure and the loop ``matrices'' displayed in the Supplemental Material clearly demonstrate that the model based on the SW particles does not provide a satisfying agreement with experiment for any possible combination of its main parameters (anisotropy and exchange constants). The main problem is the combination of the relatively low coercivity (6~kOe) and the very slow approach to saturation in the measured hysteresis loop, where even at 50~kOe the sample is not completely saturated. These features lead to the inherent contradiction in frames of the SW model (even extended by the two interparticle interactions listed above). In particular, the simulated coercivity value coincides with experiment for $q = 0.2$ (i.e., for the anisotropy being five times smaller than the bulk value of Nd$-$Fe$-$B), while the simulated magnetization behavior at large fields for this particular $q$ is completely different than the measured data. On the other hand, the slow approach to saturation at large fields could be well reproduced in simulations only by using the bulk anisotropy value ($q = 1$), while this value of $K$, in turn, resulted in a simulated coercivity $H_{\rm c} \approx 33 \, {\rm kOe}$, which is five times larger than the measured one.

\begin{figure}[htb]
\centering
\resizebox{0.95\columnwidth}{!}{\includegraphics{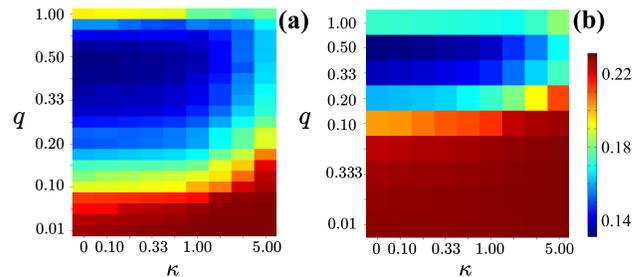}}
\caption{Difference $\Delta$ between experiment and simulation for the model based on SW particles without (a) and with (b) the magnetodipolar interaction as a function of the anisotropy-constant factor $q$ and the exchange-coupling factor $\kappa$. Note the different $q$-scales.}
\label{DeltaSWboth}
\end{figure}

The overall quality of the fit $\Delta$ computed according to Eq.~(\ref{fitcriterion}) is shown in Fig.~\ref{DeltaSWboth} as a color-coded plot $\Delta(\kappa, q)$. Results for the SW-based model both without [Fig.~\ref{DeltaSWboth}(a)] and with [Fig.~\ref{DeltaSWboth}(b)] the magnetodipolar interaction are presented. To obtain this set of results, simulations have been performed for $N_q \times N_\kappa = 21 \times 9 = 189$ parameter pairs for the model without the magnetodipolar interaction and for $N_q \times N_\kappa = 9 \times 9 = 81$ ($q, \kappa$)-pairs including this interaction. The leftmost column ($\kappa = 0$) in Fig.~\ref{DeltaSWboth}(a) corresponds to the standard SW model of non-interacting single-domain magnetic particles with uniaxial anisotropy. We note that exchange values of $\kappa > 1$ cannot correspond to a real Nd$-$Fe$-$B-based material; these values have been used solely to understand the model behavior for a very strong exchange interaction. 

Analyzing both data sets, we note, first of all, that the magnetodipolar interaction plays a significant role only when the anisotropy constant is relatively low ($q \leq 0.2$) \textit{and} the exchange constant is high $\kappa \geq 1$. This observation is naturally explained by the well known ``averaging-out'' of the random single-grain anisotropy in exchange-interacting systems (Herzer model, see Ref.~\onlinecite{HerzerModel}).  Furthermore, we emphasize that even in the parameter region with the best fit quality (blue regions in color plots), the average deviation per point $\Delta$ between measured and experimental data is unacceptably high ($\Delta > 0.14$). Taking into account that experimental errors and statistical errors of simulated results are both very small (the latter is due to the large number of grains which we could simulate), we point out that this deviation is due to a significant {\it systematic} discrepancy between measured and simulated results in the whole scanned parameter space.

Thus we conclude that the modified SW model based on single-domain grains with uniaxial anisotropy, including intergrain exchange and magnetodipolar interactions, cannot fit the experimental result in a satisfactory manner. The quality of the fit remains unacceptably poor even if we assume that (i) the anisotropy of the Nd$-$Fe$-$B nanograins may significantly deviate from the bulk value of Nd$-$Fe$-$B and (ii) the intergrain exchange may be arbitrarily weak ($0 < \kappa < 1$). We point out that the major problem of this model is the incompatibility of the very slow approach-to-saturation behavior on one side (which could only be fitted using very high anisotropy values close to the bulk value of Nd$-$Fe$-$B) and the relatively low experimental coercivity $H_{\mathrm{c}} = 6.1 \, \mathrm{kOe}$ on the other side (which can be fitted only using a relatively low anisotropy of $K \approx H_{\mathrm{c}} M_{\mathrm{S}} \approx 7.9 \times 10^6 \, \mathrm{erg/cm^3}$). Here we also note that the coercivity estimation using the SW model with standard Nd$-$Fe$-$B parameters ($M_{\mathrm{S}} = 1300 \, \mathrm{G}$ and $K_{\mathrm{bulk}} = 4.3 \times 10^7 \, \mathrm{erg/cm^3}$) results in $H_{\mathrm{c}}^{\mathrm{SW}} \approx K/M_{\mathrm{S}} \approx 33 \, \mathrm{kOe}$.

\subsection{Model based on core-shell particles}

With the aim to achieve a better understanding of the magnetization reversal of the $\mathrm{Nd}_2\mathrm{Fe}_{14}\mathrm{B}/\alpha$-$\mathrm{Fe}$ nanocomposite and to resolve the contradiction described in the previous section, we have implemented a core-shell particle model for the description of the Nd$-$Fe$-$B grains. In this model each $20 \, \mathrm{nm}$-sized grain is supposed to consist of a magnetically hard core with parameters as for bulk Nd$-$Fe$-$B, which is surrounded by a shell with different magnetic parameters. 
\begin{figure}[htb]
\centering
\resizebox{0.85\columnwidth}{!}{\includegraphics{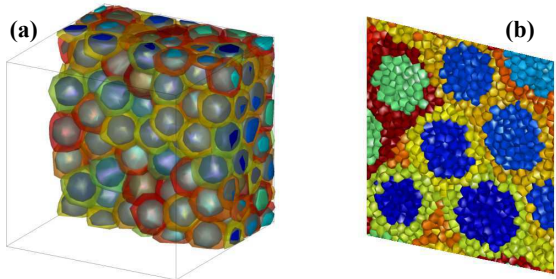}}
\caption{(a) Vertical cut through the core-shell microstructure used in the micromagnetic simulations of the Nd$-$Fe$-$B nanocomposite. (b) Example of a particular mesh-element distribution; the typical mesh-element size is $2 \, \mathrm{nm}$. Cold (warm) colors represent, respectively, cores (shell) regions.}
\label{coreshell3D}
\end{figure}
The core-shell model should take into account changes in the magnetic characteristics of the Nd$-$Fe$-$B crystallites near their surface, which may be imperfect due to the melt-spinning manufacturing process. A typical core-shell microstructure used in our simulations is shown in Fig.~\ref{coreshell3D}(a). In order to resolve the magnetization distribution inside a shell within a standard micromagnetic paradigm, we have set the mesh-element size to $2 \, \mathrm{nm}$; a small part of the corresponding mesh structure is presented in Fig.~\ref{coreshell3D}(b).

In order to gain physical insights into the role of the system parameters characterizing the particle shells, we carried out micromagnetic simulations for different anisotropy constants and exchange-stiffness values of the shells. Specifically, we have introduced dimensionless coefficients $q_{\rm s}$, $a_{\rm s}$, and $\kappa_{\rm ss}$, which describe the change of the anisotropy and exchange interactions within the shell as $K_{\rm shell} = q_{\rm s} K_{\rm bulk}$ and $A_{\rm shell} = a_{\rm s} A_{\rm bulk}$, and the exchange between different shells as $A_{\rm{shell-shell}} = \kappa_{\rm ss} A_{\rm bulk}$. For the exchange coupling between the core and its shell, the bulk value of the exchange stiffness of $\mathrm{Nd}_2\mathrm{Fe}_{14}\mathrm{B}$ was chosen. We remind the reader that for all results shown below the average grain size is $d_{\rm av} = 20 \, \mathrm{nm}$ with a core volume fraction of $c_{\rm core} = 40 \, \%$, so that the shell thickness is $h_{\rm sh} \approx 2.6 \, \mathrm{nm}$, if not stated otherwise. 

\begin{figure*}[htb]
\centering
\resizebox{1.5\columnwidth}{!}{\includegraphics{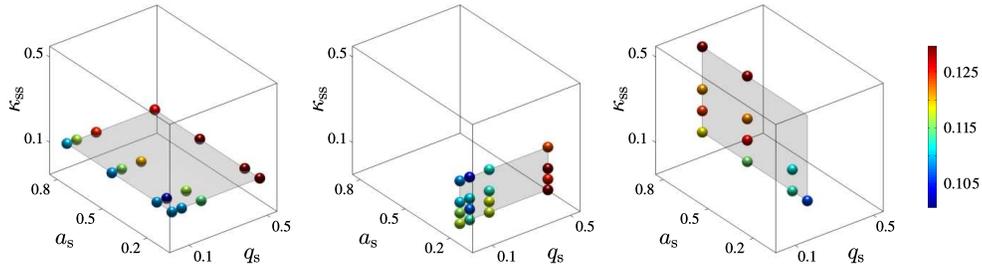}}
\caption{Parameter space used in the micromagnetic simulations of core-shell-based nanocomposites and quality-of-fit parameter $\Delta$ (shown as color circles) for various shell parameters.}
\label{paramspace}
\end{figure*}

\begin{figure}[htb]
\centering
\resizebox{0.95\columnwidth}{!}{\includegraphics{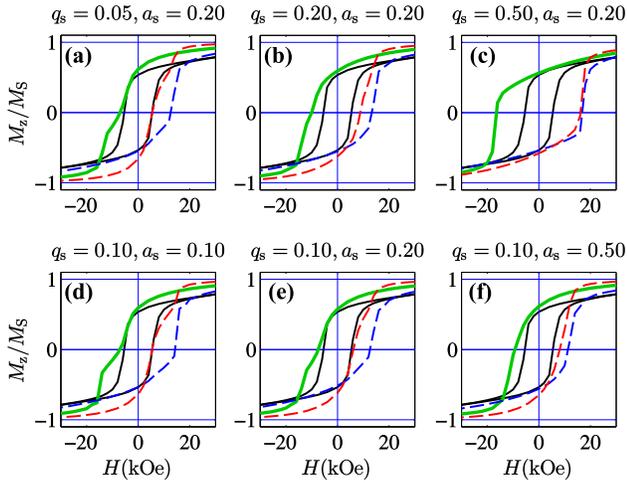}}
\caption{Upper part of simulated hysteresis loops (green) obtained by the core-shell model without the magnetodipolar interaction for different shell anisotropy constants $q_{\rm s}$ (a)$-$(c) and different intrashell exchange constants $a_{\rm s}$ (d)$-$(f) ($\kappa_{\rm ss} = 0.1$ for all loops). Lower parts of the core (blue dashed lines) and shell (red dashed lines) hysteresis loops are also shown. The experimentally measured loop is drawn in black.}
\label{cs_qs-as_selected}
\end{figure}

\begin{figure}[htb]
\centering
\resizebox{0.95\columnwidth}{!}{\includegraphics{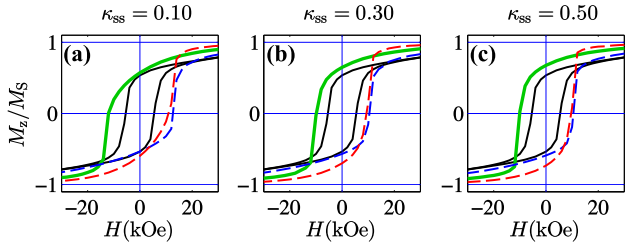}}
\caption{Same as Fig.~\ref{cs_qs-as_selected}, but for different  exchange coupling constants $\kappa_{\rm ss}$ between the shells ($a_{\rm s} = 0.8, q_{\rm s} = 0.2$).}
\label{cs_as-kappass_selected}
\end{figure}

Figure~\ref{paramspace} depicts an overview of the parameter space used in our simulations. Three large sets of micromagnetic modeling corresponding to ``plane cuts'' in this parameter space were carried out: (i) $\{ q_{\rm s} - a_{\rm s} \}$-plane at $\kappa_{\rm ss} = 0.1$; (ii) $\{ q_{\rm s} - \kappa_{\rm ss} \}$-plane at $a_{\rm s} = 0.2$; (iii) $\{ a_{\rm s} - \kappa_{\rm ss} \}$-plane at $q_{\rm s} = 0.2$. Points with $\kappa_{\rm ss} > a_{\rm s}$ have been removed from the diagram, because for physical reasons the intergrain exchange can not be larger than the exchange within the grain shell. The magnetodipolar interaction was neglected in these calculations; its role will be discussed separately.
Selected results from these three sets are presented in Figs.~\ref{cs_qs-as_selected} and \ref{cs_as-kappass_selected}, where the complete experimental loops (black) and upper parts of simulated loops (green) are plotted. Additionally---and this is a great advantage of micromagnetic modeling---we can extract and plot separately the normalized magnetization reversal curves of all cores (blue dashed lines) and of all shells (red dashed lines). This possibility allows us to analyze the behavior of these constituents of the microstructure separately. In order not to overload the figures, only the lower parts of these core and shell loops are shown in Figs.~\ref{cs_qs-as_selected} and \ref{cs_as-kappass_selected}. Complete ``matrices'' of hysteresis loops for all parameter points shown in Fig.~\ref{paramspace} are contained in the Supplemental Material. Based on these results, we are able to analyze separately the influence of each shell parameter which was varied in simulations.

For low values of the anisotropy $q_{\rm s}$ and intrashell exchange $a_{\rm s}$ constants (so that $\kappa_{\rm ss} < a_{\rm s}$ is also small), the simulated loops exhibit a two-phase behavior with a kink at about $-15 \, \mathrm{kOe}$, as it can be seen in Fig.~\ref{cs_qs-as_selected}(a). The reason for this behavior is evident from the inspection of partial loops for the magnetization reversal of cores (blue dashed lines) and shells (red dashed lines): due to the much lower anisotropy and weak intrashell coupling the grain shells reverse in much smaller fields than the cores, producing a kink in the total hysteresis. Increase of the shell anisotropy towards the bulk value results in the increase of the system coercivity $H_{\rm c}^{\rm cs}$ and the gradual disappearance of the two-phase behavior (see the sequence of loops (a) $\rightarrow$ (c) in Fig.~\ref{cs_qs-as_selected}). Due to the exchange coupling between the core and its shell, their reversal fields nearly coincide already for $q_{\rm s} = 0.5$. We note that for this shell anisotropy the coercivity $H_{\rm c}^{\rm cs}$ of the whole system is still considerably smaller than $H_{\rm c}^{\rm SW}$ for the SW-based model. The relation between the coercivities $H_{\rm c}^{\rm cs}/H_{\rm c}^{\rm SW}$ approximately corresponds to the relation between the effective grain anisotropy and the bulk anisotropy $K_{\rm eff}/K_{\rm bulk} = c_{\rm core}+q_{\rm s}(1-c_{\rm core})$ (we remind the reader that the directions of the anisotropy axes of the core and the shell are assumed to be the same). In addition, increasing of $q_{\rm s}$ leads to a small decrease of remanence in the studied parameter region, because various grains become more independent and the remanence tends to its ideal value of $j_{\rm R}^{(0)} = 0.5$ in the disordered system of uniaxial particles.

The increase of the exchange coupling {\it within} the shell $a_{\rm s}$ also eliminates the kink in the hysteresis loop, but leads only to a minor increase of $H_{\rm c}^{\rm cs}$ for the given value of $q_{\rm s}$ (see loops (d) $\rightarrow$ (f) in Fig.~\ref{cs_qs-as_selected}). This effect can be explained by a more coherent reversal of the shell for higher values of $a_{\rm s}$; due to the strong core-shell coupling this reversal leads to the switching of the core magnetization in much smaller fields than $H_K^{\rm core} \sim K_{\rm bulk}/M_{\rm S}$, strongly decreasing the coercivity of the whole system. 
If the exchange coupling between {\it different} shells (grains) $\kappa_{\rm ss}$ increases, then the coercivity decreases and the remanence increases (Fig.~\ref{cs_as-kappass_selected}). Both effects are a direct consequence of the more cooperative magnetization reversal of different grains for larger intergrain exchange (larger $\kappa_{\rm ss}$).
Based on the above results we conclude that the ranges of parameters where a better agreement between experiment and simulations is achieved ($a_{\rm s} = 0.1 \,...\, 0.2$; $\kappa_{\rm ss} = 0.0 \,...\, 0.2$; $q_{\rm s} = 0.05 \,...\, 0.2$) correspond to the physical picture of a nanocomposite in which the regions near grain surfaces---represented by shells in our model---have ``deteriorated'' magnetic properties in comparison to the bulk material. This deterioration could occur due to various technological reasons during the nanocomposite manufacturing.

\begin{figure}[htb]
\centering
\resizebox{0.7\columnwidth}{!}{\includegraphics{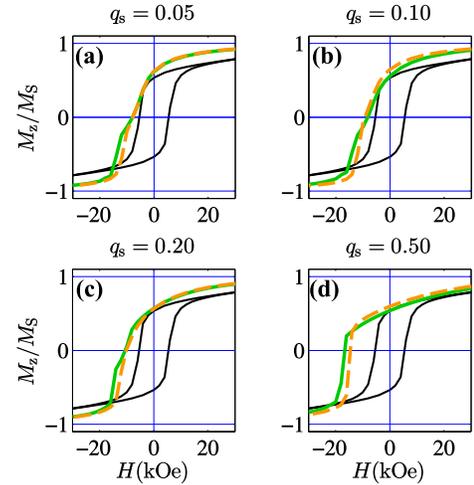}}
\caption{Comparison between core-shell systems without (green solid curve) and with (orange dashed curve) the magnetodipolar interaction for different shell anisotropy constants $q_{\rm s}$ ($a_{\rm s} = 0.2$; $\kappa_{\rm ss} = 0.05$).}
\label{cs_qs-kappass_dipcomp}
\end{figure}

The role of the magnetodipolar interaction in our core-shell model of the grain structure is demonstrated in Fig.~\ref{cs_qs-kappass_dipcomp} for the sets of magnetic parameters chosen based on the previous calculations (with $a_{\rm s} = 0.2$ held fixed). As it can be seen from the comparison of corresponding loops, the main effect of the magnetodipolar interaction is the diminishing or even complete elimination of the two-step character of the magnetization reversal of a nanocomposite [Fig.~\ref{cs_qs-kappass_dipcomp}~(a), (b) and (c)]. The most probable reason for this effect is the long-range nature of the magnetodipolar interaction.
Moreover, this interaction results in a decreased coercivity, even in systems with a relatively large shell anisotropy (see Fig.~\ref{cs_qs-kappass_dipcomp}(d)]. Similar to the increase of $\kappa_{\rm ss}$ [Fig.~\ref{cs_as-kappass_selected}), the presence of the magnetodipolar interaction also increases the intergrain coupling, resulting in a more cooperative magnetization reversal of the particle shells (``softer'' phase in our material); the reversal of the shells magnetization leads, in turn, to the magnetization switching of the whole particle in smaller external fields. This effect has already been been observed in Ref.~\onlinecite{schrefl94}, where isotropic nanocrystalline Nd$-$Fe$-$B was simulated.

\begin{figure}[htb]
\centering
\resizebox{0.99\columnwidth}{!}{\includegraphics{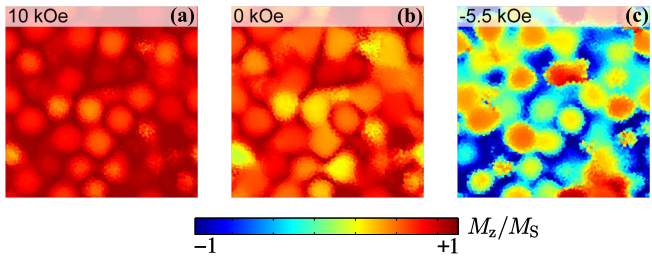}}
\caption{Magnetization distribution (two-dimensional cut out of the three-dimensional distribution) at selected points on the hysteresis curve (approach-to-saturation, remanence, coercivity) using the core-shell model ($q_{\rm s} = 0.16$, $a_{\rm s} = 0.20$, $\kappa_{\rm ss} = 0.20$, $d_{\mathrm{core}} = 14.1 \, \mathrm{nm}$).}
\label{spintexture}
\end{figure}
To establish the relation between the microstructure and the magnetization configuration in our system, we show in Fig.~\ref{spintexture} the magnetization distribution at selected points of the hysteresis curve ((a) large positive field; (b) remanence; (c) coercivity). It can be clearly seen that in high fields the shells exhibit a larger magnetization projection in the field direction than the cores, since the shell anisotropy constants are reduced compared to the core regions. This situation prevails down to the remanent state, where a qualitatively similar spin distribution is observed. However, at negative fields---see panel (c) for $H = - H_{\mathrm{c}}$---the shells reverse their magnetization ``easier'' than the cores (compare, e.g., the dashed curves in Fig.~\ref{superparaMdfine}), which again can be attributed to the reduced anisotropy in the shell region.

\begin{figure}[htb]
\centering
\resizebox{0.7\columnwidth}{!}{\includegraphics{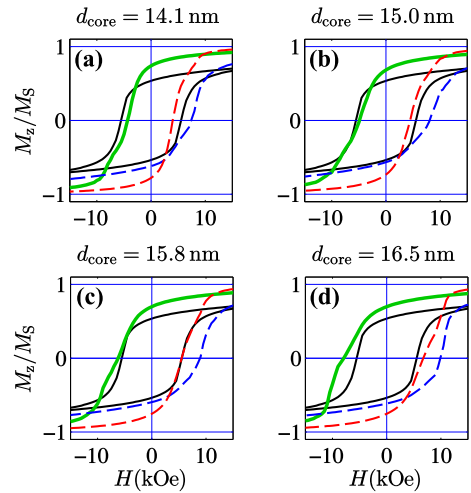}}
\caption{Hysteresis loops (green, upper part) obtained in the core-shell model with the magnetodipolar interaction for different core-volume fractions ($q_{\rm s} = 0.05$; $a_{\rm s} = 0.2$; $\kappa_{\rm ss} = 0.2$). Only the lower parts of the core (blue dashed lines) and shell (red dashed lines) hysteresis loops are shown. Experimentally measured loop is shown in black.}
\label{cs_ccore_qs}
\end{figure}

In order to complete the analysis of the core-shell-based model, we have studied the effect of its most important \textit{structural} parameter, the core volume fraction $c_{\rm core}$. Corresponding selected results are displayed in Fig.~\ref{cs_ccore_qs}, where the values of the core diameter $d_{\rm core}$ representing the increasing core volume fraction are shown. A more complete data set in form of a ``matrix'' of hysteresis loops in the ($d_{\rm core} - q_s$)-plane is presented in the Supplemental Material.
As expected, the coercivity increases with increasing volume fraction of the hard magnetic material, while the remanent magnetization remains nearly constant in the studied range of $d_{\rm core}$. In particular, for the coercivity the best agreement between simulations and experiment is found for the following parameter set: $q_{\rm s} = 0.16$, $a_{\rm s} = 0.20$, $\kappa_{\rm ss} = 0.20$, $d_{\mathrm{core}} = 14.1 \, \mathrm{nm}$. The hysteresis loop corresponding to this set of parameters is shown in Fig.~\ref{superparaMdfine}(b) in green. We emphasize that while the $H_{\mathrm{c}}$ is perfectly fitted, the high-field behavior is not reproduced adequately [although the agreement for this behavior is much better than for the SW-based model, see Fig.~\ref{HystSWwithdipoleKAPPA033}(b)]. Hence, we come to the conclusion that even the much more sophisticated model based on the core-shell structure of Nd$-$Fe$-$B grains cannot satisfactorily explain the measured hysteresis loop for the Nd$-$Fe$-$B nanocomposite under study.

\subsection{Core-shell model including superparamagnetic clusters}

A possible explanation for the deviation between simulation and experiment, in particular, in the approach-to-saturation regime, might be the existence of small clusters in the superparamagnetic state. In order to achieve a better understanding of the magnetization reversal processes of Nd$-$Fe$-$B composites consisting of nanosized Nd$-$Fe$-$B grains, we have included into our model a contribution to the system magnetization arising from magnetic clusters within the sample which are so small that they would exhibit a superparamagnetic behavior at room temperature. Such clusters could e.g.\ form in the intergrain regions of a nanocomposite. In standard coarse-grained magnetic materials the corresponding contribution is negligibly small, because the surface-to-volume ratio for relatively large grains (with a grain size larger than about 1 $\mu {\rm m}$) is very small. However, for composites with grains in the nm-regime the contribution from small clusters, formed in the interface regions between ferromagnetic grains, could be substantial. In our system such clusters can be formed preferably from the $\alpha$-Fe phase, but can also be represented by very small ($<$ 1~nm) Nd$-$Fe$-$B particles.

It is commonly known that the field dependence of the magnetization of a system of superparamagnetic particles is given by the Langevin function $L(x) = {\mathrm{coth}}(x) - 1/x$, where $x = \mu H/kT$, with $\mu$ being the particle magnetic moment. Taking into account this contribution, we have fitted the normalized total experimental magnetization loop as the sum of ferro- and superparamagnetic terms: 
\begin{equation}
\label{langevinfitnew}
M_{\rm tot}^{\rm exp} = 
\frac{c_{\mathrm{fm}} \, M^{\mathrm{cs}}_{\mathrm{s}} m^{\mathrm{cs}}(H) + 
(1 - c_{\mathrm{fm}}) \, M^{\mathrm{Fe}}_{\mathrm{s}} L(H/H_{\mathrm{L}})}
{c_{\mathrm{fm}} \, M^{\mathrm{cs}}_{\mathrm{s}} + (1 - c_{\mathrm{fm}}) \, M^{\mathrm{Fe}}_{\mathrm{s}}}, 
\end{equation}
Here, $c_{\rm fm}$ denotes the volume fraction of the ferromagnetic material in the sample, $m^{\rm cs}(H)$ represents the normalized hysteresis loop of the Nd$-$Fe$-$B core-shell particles and $M^{\rm cs}_{\rm s} = 1281 \, \rm{G}$ is the saturation magnetization of these particles. In the superparamagnetic term, $H_{\rm L} = kT/\mu$ is the characteristic Langevin field and $M^{\rm Fe}_{\rm s} = 1750 \, \rm{G}$ is  the saturation magnetization of Fe (we assume that superparamagnetic particles consist of Fe atoms). For the ferromagnetic contribution $m^{\rm cs}(H)$ we have used the hysteresis loop of the core-shell model which provided the best fit of the coercivity $H_{\rm c}$ as explained at the end of the previous subsection. The Langevin field $H_{\rm L}$ and the volume fraction $c_{\rm fm}$ of the ferromagnetic material are the fitting parameters. 

\begin{figure}[htb]
\centering
\resizebox{0.90\columnwidth}{!}{\includegraphics{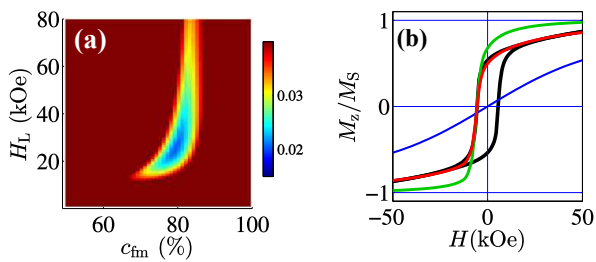}}
\caption{(a) Difference $\Delta$ between the measured and simulated hysteresis curve computed using Eq.~(\ref{langevinfitnew}) as a function of the volume fraction $c_{\mathrm{fm}}$ and the Langevin field $H_{\rm L}$; the best-fit loop from the pure core-shell model [green curve in panel (b)] was used as the term $m^{\rm cs}(H)$ in Eq.~(\ref{langevinfitnew}). (b) Comparison between the best-fit loop (red) [based on Eq.~(\ref{langevinfitnew})] and the experimental data (black). The Langevin contribution is drawn as the blue curve.}
\label{superparaMdfine}
\end{figure}

The result of this fitting procedure is presented by the two-dimensional color-coded plot in Fig.~\ref{superparaMdfine}(a) and by the red curve in panel (b) of the same figure. The best-fit magnetization curve shows an excellent agreement with experiment. The volume fraction of superparamagnetic particles extracted from this fit is $\approx 20 \, \%$. From the obtained Langevin field $H_{\rm L} = 25 \, \rm{kOe}$ and using $T = 300 \, \mathrm{K}$, we have estimated the diameter of the Fe particles (assuming a spherical shape) as $1.2 \, \rm{nm}$, which represents a reasonable value for superparamagnetic clusters consisting of $\alpha$-Fe. 

The estimated best-fit value of $20 \, \%$ for the volume fraction of superparamagnetic Fe clusters appears to be too high as compared to the nominal $5 \, \mathrm{wt.~\%}$ Fe content of the experimental Nd$-$Fe$-$B sample ($5 \, \mathrm{wt.~\%}$ correspond to $4.7 \, \mathrm{vol.~\%}$ assuming a mass density of $7.4 \, \mathrm{g/cm^3}$ for $\mathrm{Nd}_2\mathrm{Fe}_{14}\mathrm{B}$). There are several possible explanations for this mismatch: (i) In the fitting procedure, we have not taken into account a possible size distribution of superparamagnetic clusters. (ii) The superparamagnetic contribution was introduced phenomenologically via Eq.~(\ref{langevinfitnew}). A more rigorous implementation of this feature into the micromagnetic code would require an additional step in the energy minimization algorithm: namely, the local dipolar fields from ferromagnetic grains affect the magnetization of superparamagnetic clusters and vice versa, so that this interaction should be taken into account via an iterative convergence procedure; the corresponding implementation is in progress. (iii) Besides the possible superparamagnetic Fe clusters, other weakly magnetic or nonmagnetic phases may have formed during the crystallization process of the amorphous melt-spun precursor material \cite{suzukism2000}.

\begin{figure}[htb]
\centering
\resizebox{0.90\columnwidth}{!}{\includegraphics{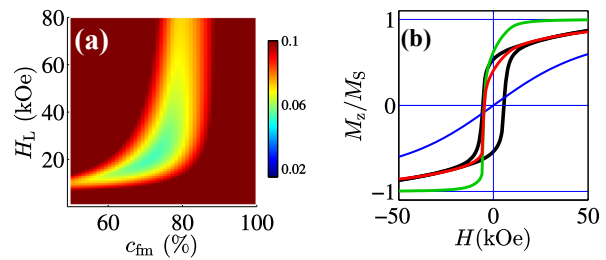}}
\caption{Same as Fig.~\ref{superparaMdfine}, with the ferromagnetic contribution described by the best-fit SW model ($q = 0.20$, $\kappa = 0$).}
\label{superparaMswfine}
\end{figure}

We emphasize that the same procedure applied for the model based on SW particles yields, first, a large systematic difference between the simulated and measured loop in both positive and negative fields with $\vert H \vert \leq H_{\rm c}$ [see Fig.~\ref{superparaMswfine}(b)] and, second, a much larger volume fraction of superparamagnetic particles ($\approx 30 \, \%$). This result demonstrates that the SW-based model is not at all suitable for the description of our system even when it is complemented by a superparamagnetic contribution.

If the above postulated superparamagnetic contribution corresponds to physical reality, then one would expect this part of the system magnetization to scale with temperature. This question can be experimentally investigated, e.g., using AC/DC magnetometry. However, the assumed superparamagnetic effect might not be unambiguously disentangled from the dominant ferromagnetic contribution, because the magnetic parameters of Nd$-$Fe$-$B itself significantly depend on temperature, and these dependencies for the nanocrystalline material are not known with the desired accuracy. Moreover, the straightforward inclusion of the temperature dependence of the superparamagnetic phase magnetization into the micromagnetic algorithm is nontrivial for the following reason: the magnetization of a particular superparamagnetic cluster strongly depends on the local dipolar field, which arises from the magnetizations of both the ferromagnetic and the superparamagnetic phases. This nonlocal interaction should be implemented in the total-energy minimization procedure as an additional iterative procedure. Work in this direction is currently in progress.

\section{Conclusion}

We have carried out micromagnetic simulations of the magnetization reversal of a nanocrystalline Nd$-$Fe$-$B based material and compared the simulation results to experimental magnetization data obtained on a $\rm{Nd}_2\rm{Fe}_{14}\rm{B}/\alpha$-$\rm{Fe}$ nanocomposite (containing $5 \, \mathrm{wt.~\%}$ of $\alpha$-Fe). It was shown that a model based on Stoner-Wohlfarth particles cannot account for the hysteresis curve, even when the magnetodipolar interaction and the exchange coupling between the particles are taken into account. A better agreement between simulation and experiment could be achieved by employing a core-shell model of Nd$-$Fe$-$B grains, but significant systematic discrepancies between simulations and experiment were still present. Inclusion of a superparamagnetic contribution originating from very small clusters, which accounts for the high-field magnetization behavior, allowed us to achieve a very good agreement between the simulated and measured hysteresis loops. A particular strength of the presented micromagnetic approach is the possibility to systematically scan the multidimensional parameter space required to describe a realistic microstructure of a nanocomposite material, involving sample regions where the magnetic parameters are different from those of the bulk material.

\section*{Acknowledgements}

The authors thank Ivan Titov and Raoul Weber for carrying out the magnetization measurements. A.M. thanks the National Research Fund of Luxembourg for financial support (Project No.~INTER/DFG/12/07). S.E. and D.B. acknowledge financial support from the Deutsche Forschungsgemeinschaft (Project No.~BE~2464/10-3). This paper is based on results obtained from the future pioneering program ``Development of magnetic material technology for high-efficiency motors'' commissioned by the New Energy and Industrial Technology Development Organization (NEDO).


\newpage
\section*{Supplemental Material}

\begin{figure*}[htb]
\centering
\resizebox{1.60\columnwidth}{!}{\includegraphics{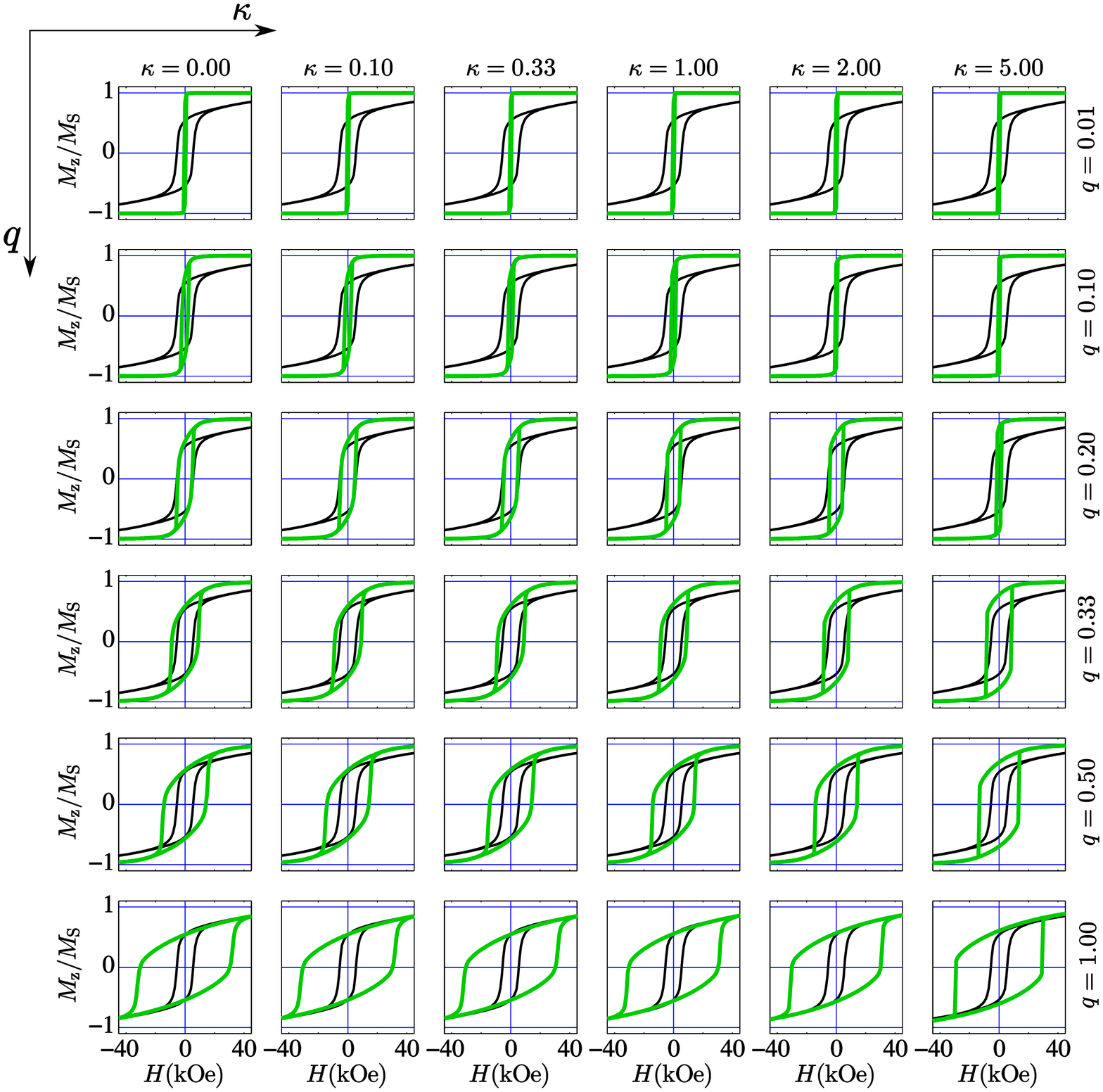}}
\caption{Simulated hysteresis loops (green) obtained for the model based on Stoner-Wohlfarth particles with the magnetodipolar interaction for different anisotropy constants $q$ and exchange constants $\kappa$. Black curve on each graph is the experimentally measured loop.}
\label{SMSW}
\end{figure*}

\begin{figure*}[htb]
\centering
\resizebox{1.50\columnwidth}{!}{\includegraphics{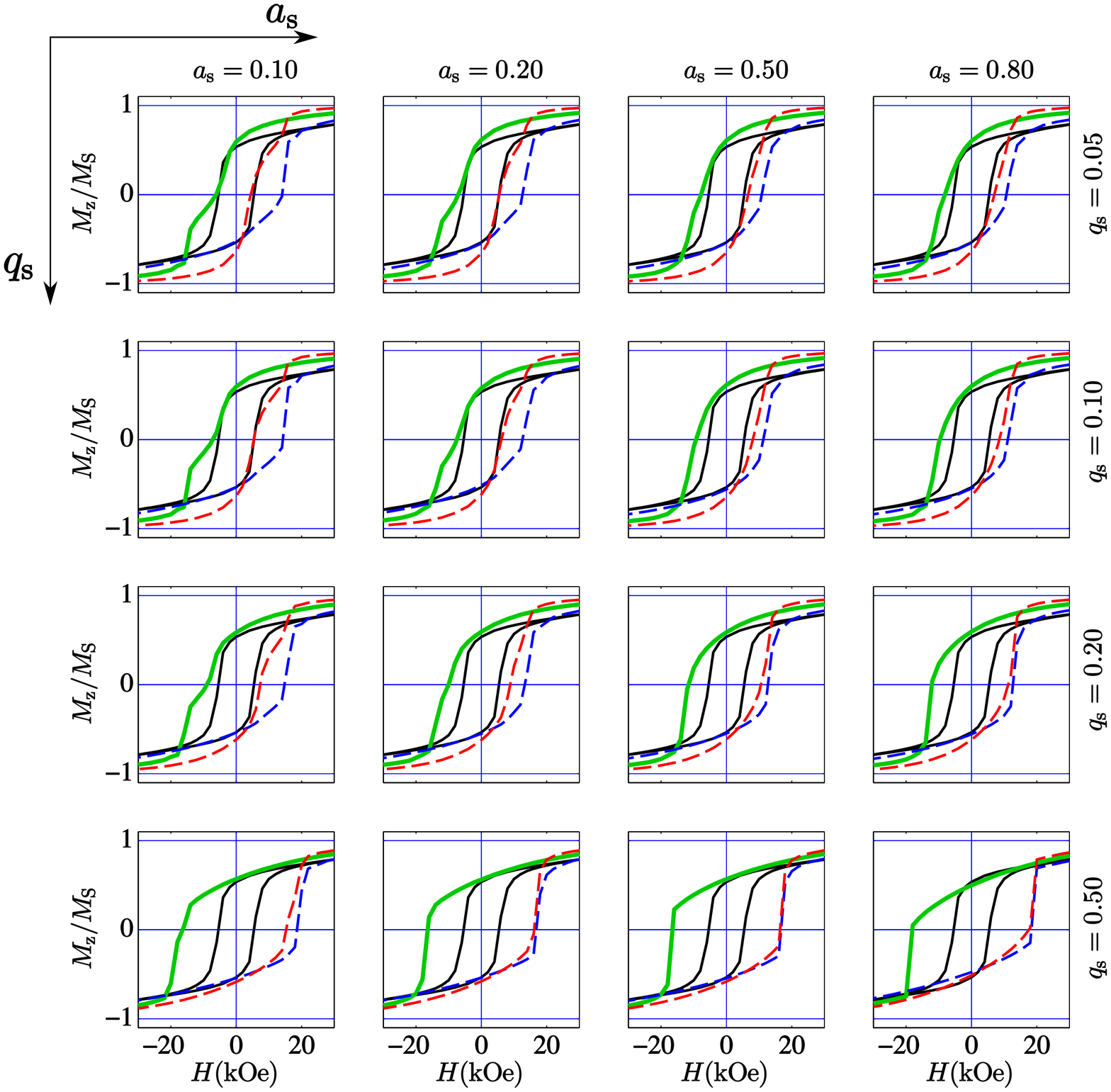}}
\caption{Upper part of simulated hysteresis loops (green) obtained by the core-shell model without the magnetodipolar interaction for different shell anisotropy constants $q_{\rm s}$ and different intrashell exchange constants $a_{\rm s}$ ($\kappa_{\rm ss} = 0.1$ for all loops). Lower parts of the core (blue dashed lines) and shell (red dashed lines) hysteresis loops are also shown. The experimentally measured loop is drawn in black.}
\label{SMcs_qs-as}
\end{figure*}

\begin{figure*}[htb]
\centering
\resizebox{1.20\columnwidth}{!}{\includegraphics{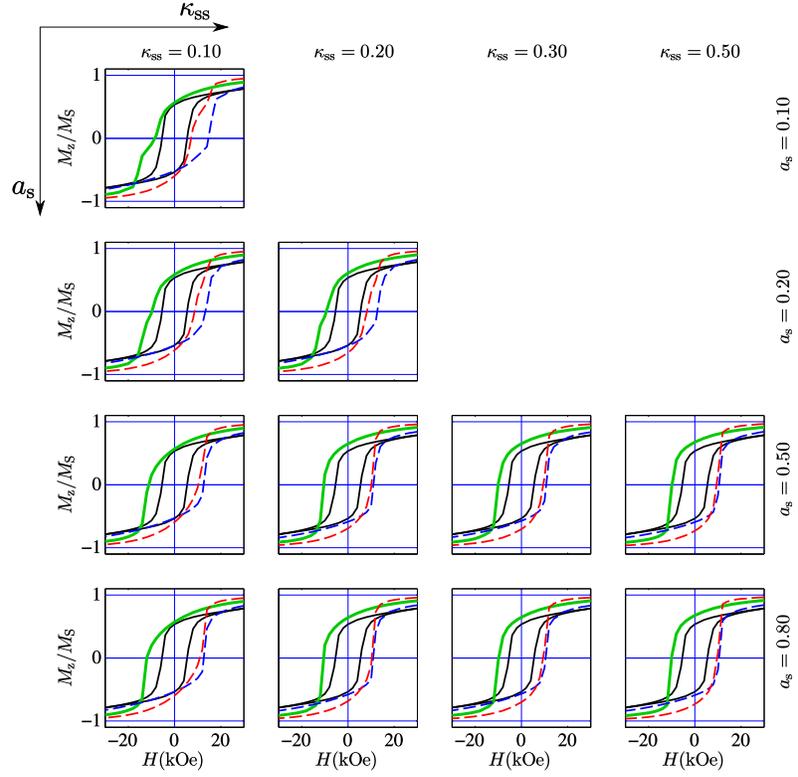}}
\caption{Same as Fig.~\ref{SMcs_qs-as}, but for different intrashell exchange constants $a_{\rm s}$ and exchange constants $\kappa_{\rm ss} \leq a_{\rm s}$ between the shells ($q_{\rm s} = 0.2$ for all loops).}
\label{SMcs_as-kappass}
\end{figure*}

\begin{figure*}[htb]
\centering
\resizebox{1.20\columnwidth}{!}{\includegraphics{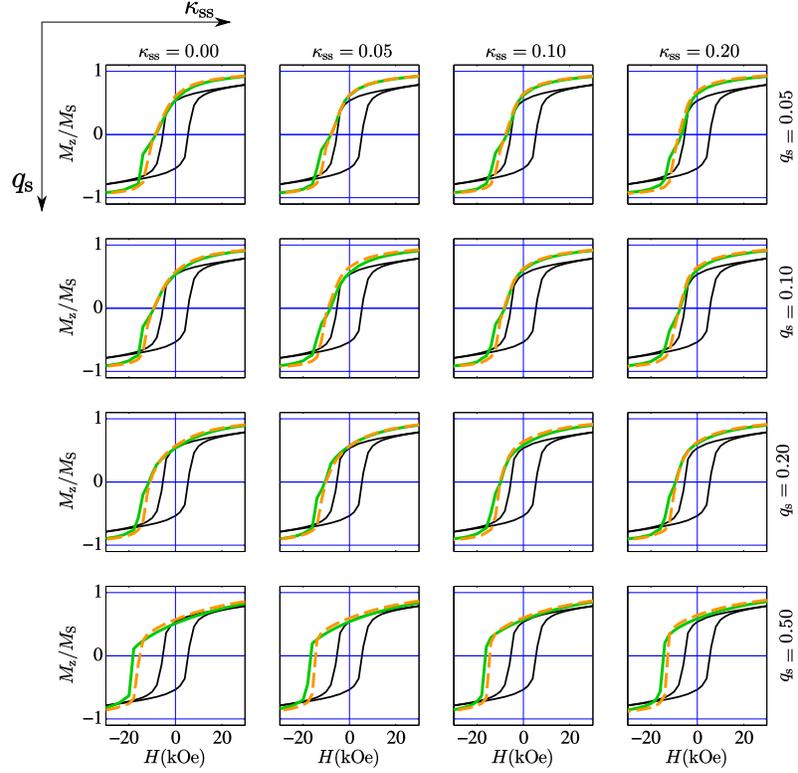}}
\caption{Same as Fig.~\ref{SMcs_qs-as}, but for different shell anisotropy constants $q_{\rm s}$ and exchange-coupling constants $\kappa_{\rm ss}$ between the shells ($a_{\rm s} = 0.2$ for all loops). Comparison between core-shell systems without (green solid curve) and with (orange dashed curve) the magnetodipolar interaction.}
\label{SMcs_qs-kappass}
\end{figure*}

\begin{figure*}[htb]
\centering
\resizebox{1.20\columnwidth}{!}{\includegraphics{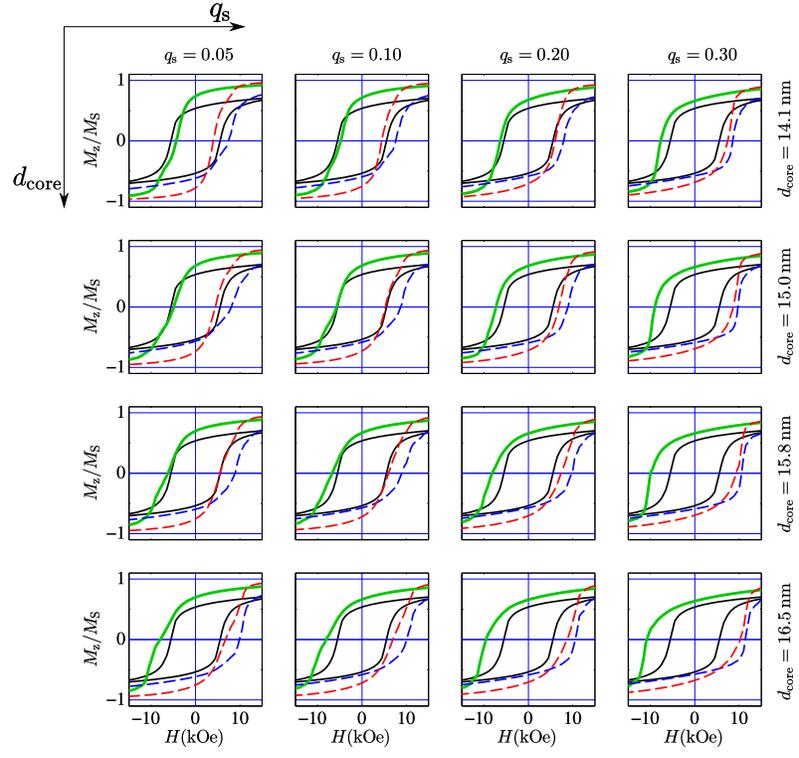}}
\caption{Same as Fig.~\ref{SMcs_qs-as}, but with the magnetodipolar interaction and for different core diameters $d_{\rm core}$ and shell anisotropy constants $q_{\rm s}$ ($a_{\rm s} = 0.2$; $\kappa_{\rm ss} = 0.2$).}
\label{SMcs_dcore-qs}
\end{figure*}

\end{document}